\documentclass[pre,twocolumn,showpacs,floatfix,superscriptaddress,aps]{revtex4}

\usepackage{graphicx}

\usepackage{amsmath,amssymb}

\begin{document}

\title{Low dimensional behavior in three-dimensional coupled map lattices}

\author{Paulsamy Muruganandam}

\affiliation{School of Physics, Bharathidasan University,
Palkalaiperur, Tiruchirappalli -- 620024, India}

\author{Gerson Francisco}

\author{Marcio de Menezes}

\affiliation{Instituto de F\'{\i}sica Te\'orica, Universidade
Estadual Paulista, R. Pamplona 145, 01405-000 S\~ao Paulo, Brazil}

\author{Fernando F. Ferreira}

\affiliation{Grupo Interdisciplinar de F\'{\i}sica da Informa\c{c}\~ao e Economia (GRIFE), Escola de Arte, Ci\^encias e Humanidades, Universidade de S\~ao Paulo,  Av. Arlindo Bettio 1000, 03828-000 S\~ao Paulo, Brazil}

\begin{abstract}

The analysis of one-, two-, and three-dimensional coupled map lattices is here developed under a statistical and dynamical perspective. We show that the three-dimensional CML exhibits low dimensional behavior with long range correlation and the power spectrum follows $1/f$ noise. This approach leads to an integrated understanding of the most important properties of these universal models of spatiotemporal chaos. We perform a complete time series analysis of the model and investigate the dependence of the signal properties by change of dimension.

\end{abstract}


\maketitle

\section{Introduction}

The widespread interest in coupled map lattices (CML) in research and applications is mainly due to their role as a family of systems  possessing universal behavior with explicit form of the local interactions. Applications of CML are found in several areas such as dynamics, turbulence, phase transitions, geophysics, optics, genetics, human information processes, and so forth \cite{Grassberger1991, Cross1993, Marcq1996, Brewster1997, Lachaux1997, Yanagita1995, Colovas1997, Cuche1997, Keeling1997, Hilgers1999, Lind2002}. It is worth to mention that Kaneko has studied spatiotemporal pattern dynamics in both 1D and 2D CML \cite{Kaneko1989}. In particular, Kaneko has shown the existence of similar patterns of different varieties in 1D and 2D CML in the weak coupling regime. However, in the strong coupling regime, the dynamics in 1D and 2D lattices differ \cite{Kaneko1989}. These coupled map lattices also exhibit the size instability wherein the dynamics is completely dependent on system size \cite{Bohr1989, Palaniyandi2005}. For example, 1D CML shows stable synchronous state at small number of lattices while it transits to spatiotemporal chaos at large system size.

Although in many instances the focus is on one- and two- dimensional lattices, in this paper we will obtain a wide range of properties that characterize the time series of 1D, 2D and 3D coupled map lattices. In this sense a comparison can be achieved on the effect of the behavior of the system under change of dimension. Due to the complexity of this endeavor we will concentrate on the logistic maps but the numerical work can be extended, using more powerful computers, to Lorenz-type and other models.

As mentioned above, the dynamical dependence on the control parameters, particularly in 1D and 2D coupled map lattices is well known \cite{Kaneko1989, Chate1992, Chate1992a} and we fix values that allow interesting chaotic regime. In the present study, we deal specifically with the time series generated by the evolution and explore their statistical, dynamic and geometric properties; we also obtain measures of dimension, and predictability when the dimension of the lattice is increased. The robustness of our conclusions was tested by taking several random points on the lattice, for each dimension, and analyzing the corresponding time series. In spite of the fact that the properties of the maps are known, subtle interrelation of the many modes of evolution will occur that are not present in 1D lattices.

Several systems of practical interest exhibit $1/f$ scaling behavior \cite{Pilgram1998, Rikvold2003} and in this paper we implement detrended fluctuation analysis (DFA) to analyze such property \cite{Peng1992, Peng1994, Hu2001}. We use statistical and other characterizations of the CML in order establish a set of properties that appear in the time series. Measures of dimension and other diagnostics will be used to detect differences of behavior of these models.

In this contribution we aim to provide several arguments that serve to compare and clarify the relationship between the time series of 1D, 2D and 3D CMLs. A new feature that emerges in 3D models, not present in lower dimensions, occurs in the plot of the root mean square of the fluctuation. Another is the increase of persistence when the dimension is increased.

The paper is organized as follows. In Section \ref{sec2} we describe how the model and its numerical implementation is built. In Section \ref{sec3} we perform the statistical analysis, which include DFA and power spectral density. The dynamics part of the paper, which includes local dimension analysis, embedding dimension, stationarity and predictability, is discussed in Section \ref{sec4}. In the last section we present a summary of results and conclusions.

\section{Description of the model}
\label{sec2}

\begin{figure}[!ht]
\begin{center}
\includegraphics[width=0.95\linewidth]{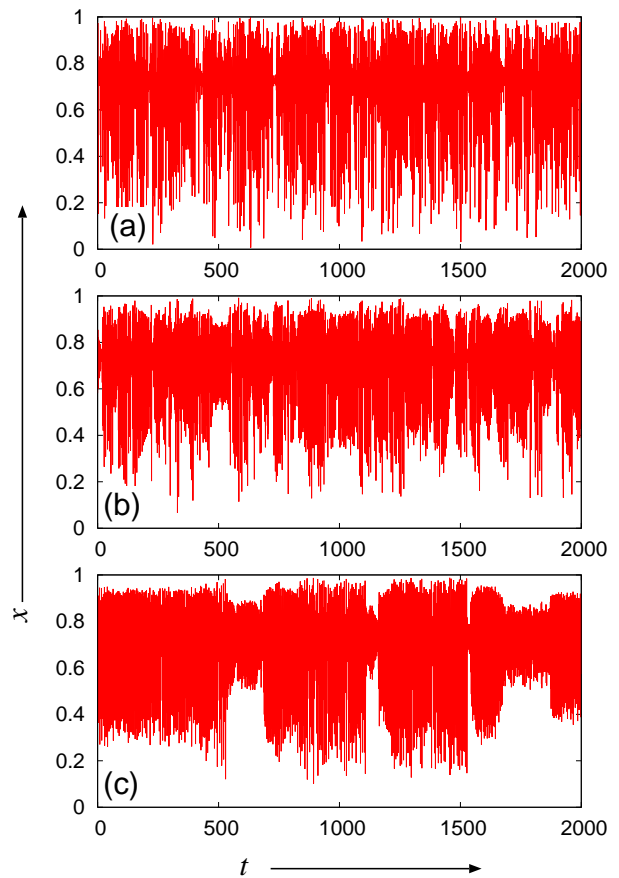}
\end{center}
\caption{(Color online) Time series of a randomly chosen lattice point for (a) one-, (b) two- and (c) three dimensional CML}
\label{fig:ts}
\end{figure}

Let us first consider an one dimensional coupled map lattices with
nearest neighbor coupling defined as~\cite{Bohr1989, Kaneko1989, Palaniyandi2005},
\begin{align}
x_{t+1}^i = (1-\epsilon) f(x_t^i) + \frac{\epsilon}{2}\left[
f(x_t^{i-1})+f(x_t^{i+1}) \right], \label{eq:cml1d}
\end{align}
where $i = 1,2,\ldots, N$ are the lattice sites and $N$ is the
number of lattice points. Each of the lattice points in
Eq.~(\ref{eq:cml1d}) is represented by the logistic map,
\begin{align}
f(x) = \mu x (1-x),\;\;\; x \in (0,1), \;\;\; \mu \in (0,4).  \label{eq:logis}
\end{align}
In the two dimensional lattices, each map is coupled to four of its
nearest neighbors and given by the following form:
\begin{align}
x_{t+1}^{i,j} = & \, (1-\epsilon) f(x_t^{i,j}) +
\frac{\epsilon}{4}\left[f(x_t^{i-1,j})+f(x_t^{i+1,j}) \right. \notag \\
& \, \left. +f(x_t^{i,j-1})+f(x_t^{i,j+1})\right]. \label{eq:cml2d}
\end{align}
One can also think of a coupled map lattices in three dimensions where each of the individual map is coupled to six of its neighbors and represented by the following set of coupled equations.
\begin{align}
x_{t+1}^{i,j,k} = & \, (1-\epsilon) f(x_t^{i,j,k}) +
\frac{\epsilon}{6}\left[f(x_t^{i-1,j,k})+f(x_t^{i+1,j,k}) \right. \notag \\
& \, \left. +f(x_t^{i,j-1,k})+f(x_t^{i,j+1,k}) 
\right. \notag \\ & \, \left.
+f(x_t^{i,j,k-1})+f(x_t^{i,j,k+1})\right].
\label{eq:cml3d}
\end{align}
For the present study we fix the parameters as $\mu = 4$, $\epsilon = 0.4$ and the size as $N=50$. The time evolution generated by these maps is shown in Fig.~\ref{fig:ts}. The 
distinguishing feature of the 3D case, Fig.~\ref{fig:ts}(c) is the variation of the amplitude of the motion resulting in the clustering of variances.

\section{Statistical analysis}
\label{sec3}

\subsection{Detrended fluctuation analysis}

Consider a time series $x(i)$, $i = 1, 2, \ldots, N$ and define an
integrated the time series $y(i)$ as,
\begin{align}
y(i) = \sum_{j = 1}^{i} \left[x(j) - \bar x\right]
\end{align}
where $\bar x$ is the mean given by
\begin{align}
\bar x = \frac{1}{N}\sum_{i = 1}^{N} x(i).
\end{align}
Then the integrated time series $y(i)$ is divided into $l$ boxes of equal size, say, $n$ where $l = \mbox{Int}\, (N/n)$. The data in each box is fitted with a linear regression function $y_{\mbox{\small fit}}(i)$, the local trend. Now detrend the integrated time series by subtracting the local trend $y_{\mbox{\small fit}}(i)$ in each box. Then the root mean square (rms) fluctuation is calculated as
\begin{align}
F(n) = \sqrt{\frac{1}{l}\sum_{k = 0}^{l-1} \left \{
\sum_{i=kn+1}^{(k+1)n}\left[ y(i) - y_{\mbox{\small fit}}(i)
\right]^{2} \right\}}.
\end{align}
By repeating the above calculation for different box sizes the relationship between the rms fluctuation $F(n)$ and the size $n$ is established. In general, there follows a power-law relation between $F(n)$ and $n$, and $F(n)$ scales as $F(n) \sim n^{\alpha}$, where $\alpha$ is called scaling exponent or DFA exponent or sometimes referred as correlation exponent. Depending on the value of $\alpha$ one can classify the time series whether it is correlated or anti-correlated. For an uncorrelated time series $\alpha = 0.5$, for positive correlations $\alpha > 0.5$, and $\alpha < 0.5$ for
anti-correlated time series. A detailed analysis on the different trends of artificial data sets are found in Ref.~\cite{Hu2001}.

When the autocorrelation $C(n)$ separated by $n$ time units, decays as
\begin{align}
C(n) \sim n^{-\gamma}, \label{eq:correl}
\end{align}
with $0 < \gamma < 1$, the fluctuation exponent obeys the following
relation
\begin{align}
\alpha = 1 - \frac{\gamma}{2}. \label{eq:alpha_gamma}
\end{align}
Note that $\gamma > 1$ for uncorrelated data. Further the DFA exponent is related to the $1/f^{\beta}$ spectral slope as
\begin{align}
\beta = 2 \alpha - 1, \label{eq:alpha_beta}
\end{align}
and $\alpha = 1$ for a $1/f$ noise spectrum and $\alpha = 1.5$ for Brownian motion (random walk) where $\beta = 2$.

\begin{figure}[!ht]
\begin{center}
\includegraphics[width=\linewidth]{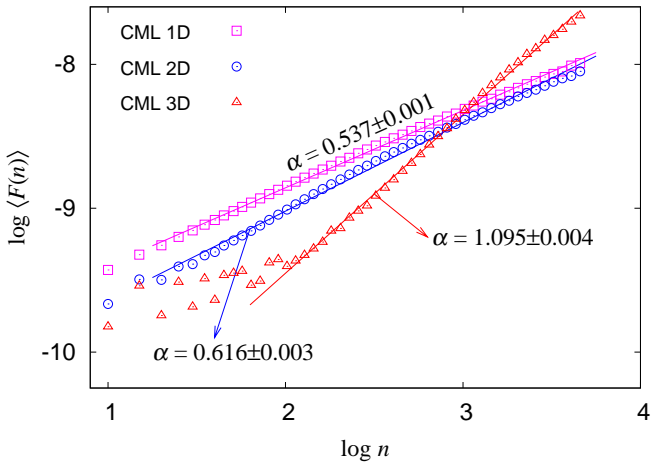}
\end{center}
\caption{(Color online) The variation of $F(n)$ vs. $n$ for the one-, two- and three-dimensional CML} \label{fig:dfa}
\end{figure}

We calculate the exponent $\alpha$ for one-, two- and three-dimensional coupled map lattices. In all the three cases $F(n)$ scales as a power law. The exponent $\alpha$ for one-dimensional CML is found to be $0.54$ which is close to $0.5$. This essentially indicates that the time series is uncorrelated. In the case of two-dimensional CML the exponent $\alpha \sim 0.62$ showing a positive correlation in the time series [see Fig.~\ref{fig:dfa}].

The time series from three-dimensional CML exhibits an entirely different behavior. In Fig.~\ref{fig:dfa} we show the variation of the fluctuation exponent, $F(n)$, as a function of $n$. It is easy to see from Fig.~\ref{fig:dfa} that there exists two regimes with a crossover, which occurs around $n \sim 10^2$. Before the crossover there is no apparent scaling present in the time series. However, for small $n$ two dominant clusterings present in the data.

For $n > 10^2$, $F(n)$ scales as $n^\alpha$ where $\alpha \approx 1$ indicating a strong positive correlation in the time series. One should note that when $\alpha = 1$, the auto-correlation exponent $\gamma = 0$ [see eq.~(\ref{eq:alpha_gamma})]. It also indicates that the time series follows $1/f^{\beta}$ noise [see eq.~(\ref{eq:psd}) below].

\subsection{Power spectral density}

Spectral density is often used to characterize many noisy time series. In the present study we use the power spectral density (PSD) analysis to find the signature of $1/f$ noise as predicted by the DFA analysis.
\begin{figure}[!ht]
\begin{center}
\includegraphics[width=0.95\linewidth]{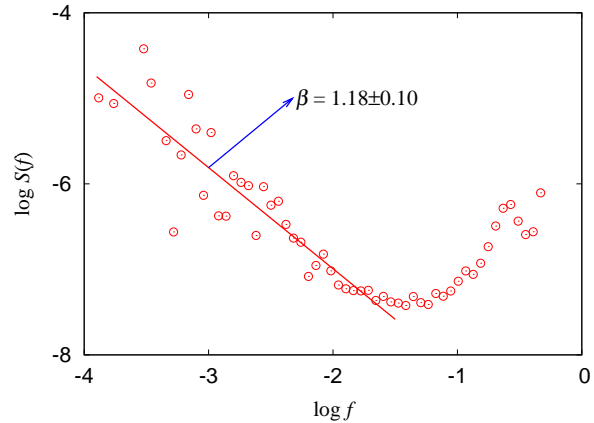}
\end{center}
\caption{(Color online) Plot of the power spectral density as a function of frequency showing $1/f^\beta$ noise spectrum for the time series obtained from 3-dimensional coupled map lattices}
\label{fig:psd_cml3}
\end{figure}
The power spectral density for a time series $\mathbf{x}$ is defined as
\begin{align}
S(f) = \vert {\mathcal F}(\mathbf{x}) \vert^2.
\end{align}
where ${\mathcal F}(\mathbf{x})$ denotes fast Fourier transform (discrete Fourier transform) of the time series $\mathbf{x}$. We are searching for a relationship of the form
\begin{align}
S(f) \sim \frac{1}{f^\beta} \label{eq:psd}
\end{align}
Fig.~\ref{fig:psd_cml3} shows the plot of the PSD as a function of frequency. Here we take the average of $S(f)$ over a range of frequencies $\Delta f$. From Fig.~\ref{fig:psd_cml3} one can see that, the PSD falls as almost $1/f^\beta$ in the case of three-dimensional CML, with ${\beta} \sim 1.18$. This is consistent with  the value of $\alpha_3$ in Fig.~\ref{fig:dfa} and eq.~(\ref{eq:alpha_beta}).

Feigenbaum \cite{Feigenbaum1983} has shown an example of deterministic dynamical systems with $1/f$ noise behavior. It has also been confirmed theoretically as a real $1/f$ spectrum density using simple quadratic map \cite{Procaccia1983}. It is a known fact that there is no generally accepted explanation for this phenomenon, in spite of its universality. However, in the context of dynamical systems, especially in chaotic systems, the intermittency is one of the sources for $1/f$ noise, and as a consequence these systems exhibit long range time and spatial correlations~\cite{Keeler1986}.

Earlier, we found that the fluctuation exponent for 1D CML is almost close to $0.5$, similar to that observed in white noise. For 2D CML the exponent increased slightly. However, there is a dramatic change in the exponent for 3D CML with $\alpha \approx 1$ [see Fig.~\ref{fig:dfa}]. This indicates the signature of $1/f$ noise. The power spectral density of the corresponding time series closely follows $1/f^{\beta}$ decay [see Fig.~\ref{fig:psd_cml3}]. The structure of correlation on data, according Fig.~\ref{fig:dfa}, clearly affected the power spectrum producing deviation around the $1/f$ curve.

\subsection{Structure function analysis}
\label{sec3c}

Structure function (SF) constitute a technique used to identify  stationarity during short time evolution. It has been successfully employed in study in turbulent flows \cite{Frisch1995} but it has a much wider applications. The SF is defined in the following way:
\begin{align}
S_q(\tau) \equiv \left\langle \left\vert Y(t_i+\tau) - Y(t_i) \right\vert^q \right\rangle, \label{eq:sf}
\end{align}
where $Y(t)$ denotes some stochastic process, or the component of some dynamical system, $i$ denotes the $i$-th data point, $\langle \cdots \rangle$ is the ensemble average and $q$ is a real number. When $q=2$ the SF is just the correlation function. If $Y(t)$ is scale invariant and self-similar (self-affine) over some time lag interval $(\tau_1 \le \tau \le \tau_2)$, then the $q$-th order structure function is expected to be fitted by
\begin{align}
S_q(\tau)=C_q \tau^{\zeta(q)},
\end{align}
Note that $C_q$ can be a function of $\tau$ whose variation is varies slower than any power of  $\tau$, and $\zeta(q)$ is the exponent of the structure function.

\begin{figure}[!ht]
\begin{center}
\includegraphics[width=0.95\linewidth]{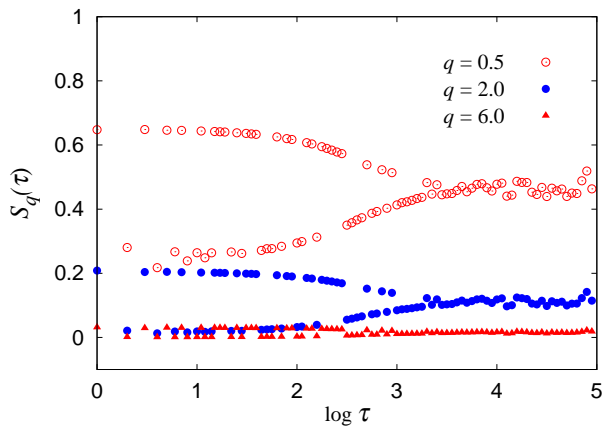}
\end{center}
\caption{(Color online) Plot of the structure function $S_q(\tau)$ as a function of delay $\tau$ for the time series obtained from 3-dimensional coupled map lattices}
\label{fig:cml3_sf}
\end{figure}

For stationary time series the exponent of the structure function is zero $\zeta(q)=0$, as a consequence of the translational invariance of any statistics defined from the data. Fig.~\ref{fig:cml3_sf} shows the SF for $q=2$. For small time lags we find two branches meaning
that a breakdown of stationarity in this region. A similar effect has been found in the study of the DFA plot in Fig.~\ref{fig:dfa}. For higher time lags the flat SF plot is indicative of stationarity in this region.

\section{Dynamical analysis}
\label{sec4}

\subsection{Local dimension}
\label{sec4a}

An interesting statistics used to characterize spatiotemporal chaotic systems is given by bred vector (BV) dimension \cite{Patil2001, Francisco2003, Muruganandam2005}. One can well analyse the local dimension of a two-dimensional spatiotemporal chaotic system using BV dimension. In the present study we apply the BV dimension analysis for the 2D and 3D CML. In the case of 3D CML we consider a two dimensional slice in order to compute the local dimension. The calculation of BV dimension is formulated as follows.

Consider a 2D spatially distributed system whose state at a given time $t_1$ is defined over a collection of points $(i,j)$. Here we take the $M-1$ nearest neighbors for each point $(i,j)$ in a square lattice with $M=25$. Logistic maps are one variable dynamical systems and in order to specify the corresponding state at a point including its neighbors we need an $M$ dimensional state vector called bred vector. Now generate $k$ distinct perturbations of the state starting at $t_0 < t_1$ obtaining $k$ local bred vectors. The $k\times k$ covariance matrix of the system is just $\mathbf{C} =
\mathbf{B^TB}$, where $\mathbf{B}$ is the $M\times k$ matrix of local bred vectors each normalized to unity. 

We order the eigenvalues of the covariance matrix as  $\lambda_1 \ge \lambda_2 \ge \ldots \ge \lambda_k$ and define the singular values of $\mathbf B$ as $\sigma_l = \sqrt{\lambda_l}$, $l = 1,2,\ldots k$. The eigenvalues $\lambda_l$ represent the amount of variance in the set of the $k$ unit bred vectors. Then the bred-vector dimension can be defined as  
\cite{Patil2001}
\begin{equation}\label{eq:bvd}
\psi_{i,j}(\sigma_1, \sigma_2,\ldots,\sigma_k) =
\displaystyle \frac{\left(\sum_{l=1}^{k}\sigma_l\right)^2}{\sum_{l=1}^{k}\sigma_l^2}.
\end{equation}
As each of the $k$ bred vectors is normalized to unity, $\psi$ assumes values in the interval $(0,k)$. In the present study, we fix the value of the number of bred vectors as $k=5$. 

We compute the local dimensions for the 2D and 3D CML for different set of initial conditions. In the case of 2D CML the local dimensions are found to lie between a minimum of $1.03$ and a maximum of $2.4$, while for the 3D CML it takes a minimum value of $\sim 1.04$ and a maximum value of $\sim 2.1$. Since this kind of dimension is related to predictability~\cite{Francisco2003, Muruganandam2005}, we conclude that the 3D CML is more predictable than the 2D case.

\subsection{Embedding dimension}
\label{sec4b}

\begin{figure}[!ht]
\begin{center}
\includegraphics[width=\linewidth]{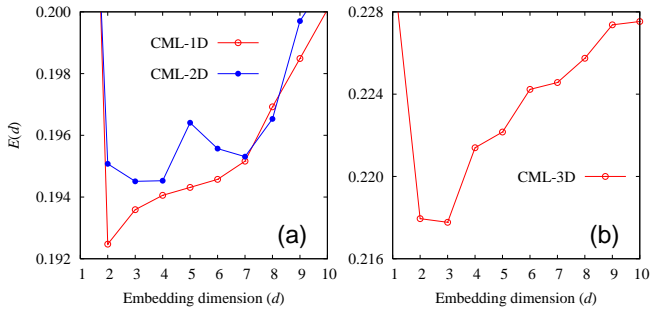}
\end{center}
\caption{(Color online) Estimation of optimal embedding dimension for the time series obtained from one-, two- and three-dimensional CMLs}
\label{fig:cao_ed}
\end{figure}

The method developed in Ref.~\cite{Cao1999} was implemented to obtain the embedding dimension of the CMLs. In our simulations this procedure was more robust and objective than the ambiguous results based on the correlation integral. The idea is to start with some initial embedding dimension $d$ and a vector $r_j$ from the set of state vectors. A simple prediction algorithm, similar to the one used in the cross prediction test of nonstationarity (see below), consists of finding the nearest neighbor $r_{\alpha(j)}$ of $r_j$  and then to evolve it one time step ahead obtaining $r_{\alpha(j)+1}$. The error in this prediction is $\parallel r_{j+1} - r_{\alpha(j)+1} \parallel$ and the error over all the set of state vectors is just the sum, $E(d)$, of each individual error. Then the embedding dimension is the value of $d$ that minimizes the function $E(d)$. This is a simple and clever way of determining the embedding dimension that does not require large data sets. [See, for example, Ref.~\cite{Eckmann1992} an estimate of the size of a time series that is necessary to give meaningful results when computing dimensions using the correlation integral. Also in Ref.~\cite{Stefanovska1997} an even more strict estimate is provided]. In Fig.~\ref{fig:cao_ed} we show
the dimension of the CMLs. We find that the 1D CML has the lowest dimension followed closely by the 2D and 3D lattices. All of them have dimensions compatible with chaotic systems.

\subsection{Stationarity and recurrence analysis}
\label{sec4c}

One of the most important tests of dynamical systems is related to stationarity. We implement three different algorithms since there is no single criterion for these tests in nonlinear dynamics. We also implement a recent method \cite{Letellier2006} using recurrence analysis to find whether these systems possess a positive maximal Lyapunov exponent.

The recurrence plot was introduced in Ref.~\cite{Eckmann1987} and it provides insights about many properties of dynamical systems in a two dimensional graphic display. The idea is to plot the times when the trajectory visits a given region in phase space. Deterministic nonlinear dynamical systems exhibit the fundamental property that states tend to return to the neighborhoods of regions already visited by the trajectory.

Consider the following function
\begin{align}
R_{ij} = \Theta(\epsilon - \parallel r_i - r_j \parallel)
\end{align}
where the state vectors $r_i$, $r_j$  are considered at discrete times with $1 \le i$, $j \le N$, $N$  the total number of states. Here $\Theta$ is the Heaviside function and $\epsilon$ a positive number whose value is discussed below.  A graph is then composed by associating a black dot in the plane with coordinates $(i, j)$ whenever the argument of the function is positive. We examine the well known baker's map \cite{Schreiber1997} for comparison and also plot the CML for dimensions 1 to 3. The values of $\epsilon$ are chosen as
\begin{figure}[!ht]
\begin{center}
\includegraphics[width=0.95\linewidth]{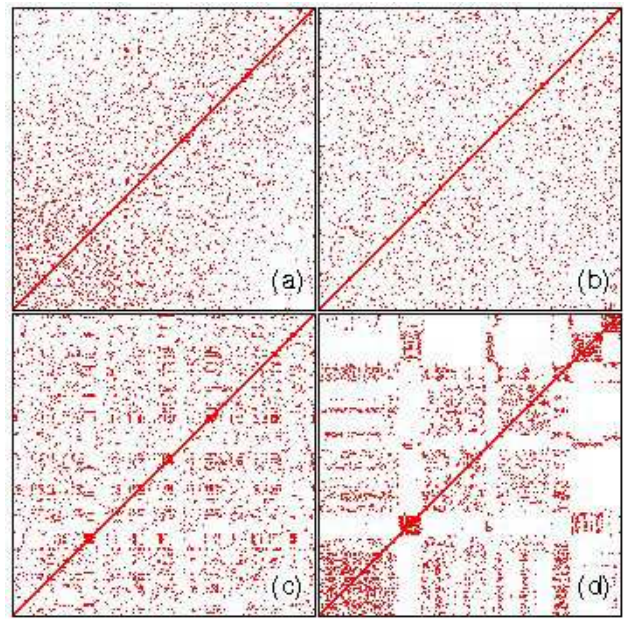}
\end{center}
\caption{(Color online) Recurrence plot for the (a) nonstationary baker's map~\cite{Schreiber1997}, (b) 1D CML, (c) 2D CML and (d) 3D CML.}
\label{fig:cml_rp}
\end{figure}
$\sqrt{d_E} \times 10\%$ of the fluctuation of the time series where $d_E$ is the embedding dimension \cite{Letellier2006}. The embedding dimension for the CMLs is 3 and the graphical analyses are not altered when this number is changed. The criterion of nonstationarity is that the density of points in the upper left and lower right corners of the recurrence plot should decrease~\cite{Marwan2003}. As we can see from the plots, we obtain inconclusive result in 3D but for the lower dimension lattices the system, according to this criterion, are stationary.

The method discussed in Ref.~\cite{Gao1999} uses the property of nonlinear dynamical systems that the recurrence time, suitably calculated, is roughly constant for stationary dynamics. The idea is to compute the recurrence time for a trajectory given an initial condition inside a ball of radius $r>0$ (more specifically, recurrence time of second type). We consider an embedding dimension 2 but the results are robust with respect to changing this parameter.  In order to eliminate the dependency on the initial condition a certain procedure is needed. This requires normalization of the recurrent time and then several initial conditions are taken. A ball of radius $r$ is centered at each initial condition  an the times for the first return to each ball are computed. The first return time is the average of these. Then the second, third, etc, return times are computed. The criterion for stationarity is that the plot of the time necessary for returns should be constant as a function of the number of recurrences. One concludes, according to  this method, that the CMLs are stationary.

Another test for stationarity \cite{Schreiber1997} requires the partition of the time series and the implementation of some predictability algorithm (or any other statistic sensitive to differences in dynamics). The algorithm is used as a criterion to evaluate how the data from subsets of the partition can be used to predict each other (thus the name cross prediction used by Schreiber \cite{Schreiber1997}). Here again the choice of embedding dimension does not affect the main conclusions and we choose it to be 3, with unit delay. Suppose the time series  $x_n$, $n = 1,2,\ldots, N$, is split into contiguous segments of size $l$ and that a statistic $\gamma_{ij}$ is constructed for each pair of segments $(i,j)$. For the present purpose we choose a simple prediction algorithm on state vectors \cite{Kantz2003} in which $\gamma_{ij}$ is represented as an error surface on the grid $(i,j)$, that is, the error of using segment $i$ to predict the dynamics in segment $j$. A plot of this surface gives an idea of whether different dynamics are at operation in different segments: when the prediction error is small then segments have similar dynamics. For nonstationary systems we would expect that the error surface has a systematic low value at the diagonal since a segment is being used to predict its own dynamics (in-sample prediction). For stationary systems the dynamics should be similar in all segments and a rough error surface is the result. From several error surfaces analyzed there is no evidence of nonstationarity in the CMLs.

Apart from the above, recurrence plots can also be used to provide information about the maximal Lyapunov exponent. It was recently shown by Letellier~\cite{Letellier2006} that recurrence plot analysis provide an estimate of the Shannon entropy and thereby indicates the existence of a positive maximal Lyapunov exponent. The key idea is to use the frequency of occurrence of diagonal segments of nonrecurrent points in the recurrence plots to estimate the Shannon entropy. In this study, we have computed the Shannon entropy using the recurrence plots for the CML and all the three cases. We find that the estimate turns out to be a positive quantity indicating the time series obtained in the 1D, 2D and 3D CML are chaotic.

\section{Summary and conclusions}
\label{sec5}

In this paper, we have made a comparative analysis on the chaotic times series generated from the one-, two- and three-dimensional coupled map lattices based on certain commonly employed tools such as fluctuation analysis, power spectral density, local dimension and recurrence quantification analysis.

We found that the plot of the root mean square of the fluctuation for the 3D lattice shows an anomalous behavior so that no scaling can be found for $n$ less than about $400$. However, there are large scaling regions for 1D and 2D maps. Further, there is notable increase in the persistence if we increase the dimension. In other words the three-dimensional CML is more persistent than one- and two-dimensional CML. Other important properties analyzed reveal that the CMLs are low dimensional and, in the 3D case, show a clear indication of $1/f$ noise. All the simulations were replicated a different points of the lattices and the conclusions are robust in this respect.

If we analyze the time series as whole, then the tests in Section \ref{sec4c} show that the system is stationary. However, it is evident from the plot of the time series in Fig.~\ref{fig:ts}(c) that there must be some change of the variance along the evolution. In order to analyze further this question we used the structure function in Section \ref{sec3c} and found that there is stationary behavior only for time values larger than about $1000$. Thus the nonsatationarity breaks the scaling for the 3D lattice dynamics for sufficiently low values of $n$. As a next step we intend to perform extensive studies on the predictability of the time series generated by the CMLs using algorithms specifically designed for low dimensional chaotic systems and also other parameter regimes of the coupled map lattices where it exhibits spatiotemporal intermittency.

\acknowledgments

The work of PM is supported by the Conselho Nacional de Pesquisa (CNPq), Brazil, the Third World Academy of Sciences, Trieste, Italy under the TWAS-UNESCO Associateship at the Centre for Excellence in the South and Department of Science and Technology (DST), Government of India. FFF thanks the Conselho Nacional de Pesquisa (CNPq), Brazil for financial support. We thank correspondence with C. Letellier on the calculation of the Shannon entropy from the recurrence plots.


\end{document}